\documentclass{PoS}

\usepackage{psfrag,epsfig,graphicx,graphics}
\usepackage{amsmath}

\newcommand{\shat}{{\hat s}}

\newcommand{\Tr}{\hbox{Tr}}

\title{QCD resummation effects in forward $J/\psi$ and very backward jet inclusive production at the LHC}

\ShortTitle{Forward $J/\psi$ and very backward jet inclusive production}

\author{\speaker{R.~Boussarie}\\
        Institute of Nuclear Physics, Polish Academy of Sciences, Radzikowskiego 152, PL-31-342 Krakow, Poland\\
        E-mail: \email{renaud.boussarie@ifj.edu.pl}}
        
\author{B.~Duclou\'e\\
Department of Physics, University of Jyv\"askyl\"a, P.O. Box 35, 40014 University of Jyv\"askyl\"a, Finland\\
Helsinki Institute of Physics, P.O. Box 64, 00014 University of Helsinki, Finland\\
Email: \email{bertrand.b.ducloue@jyu.fi}}

\author{L.~Szymanowski\\
 National Centre for Nuclear Research (NCBJ), 00-681 Warsaw, Poland\\
E-mail: \email{lech.szymanowski@ncbj.gov.pl}}
 
 \author{S.~Wallon\\
 LPT, Universit{\'e} Paris-Sud, CNRS, Universit\'e Paris-Saclay, 91405, Orsay, France {\em \&} \\
UPMC Univ. Paris 06, facult\'e de physique, 4 place Jussieu, 75252 Paris Cedex 05, France 
\\
E-mail: \email{samuel.wallon@th.u-psud.fr}}

\abstract{We propose and study the inclusive production of a forward $J/\psi$ and a very backward jet at the LHC as an observable to reveal high-energy resummation effects \`a la BFKL. Our different predictions are based on the various existing mechanisms to describe the production of the $J/\psi$, namely, NRQCD singlet and octet contributions, and the color evaporation model.
}

\FullConference{XXV International Workshop on Deep-Inelastic Scattering and Related Subjects\\
		3-7 April 2017\\
		University of Birmingham, UK}

\begin{document}


\section{Introduction}

\begin{figure}[t]
\center
\psfrag{p1}{$\hspace{-.4cm} H(p_1)$}
\psfrag{p2}{$\hspace{-.4cm} H(p_2)$}
\psfrag{q1}{$x \, p_1$}
\psfrag{q2}{$\hspace{-.2cm} x' \, p_2$}
\psfrag{k}{$k$}
\psfrag{d}{$k'$}
\psfrag{q}{$p_J$}
\psfrag{a}{$a$}
\psfrag{b}{$b$}
\psfrag{M}{$p_M$}
\psfrag{X}{$X$}
\psfrag{Y}{$Y$}
\scalebox{.9}
{
\scalebox{.82}{\raisebox{1cm}{\includegraphics[scale=1]{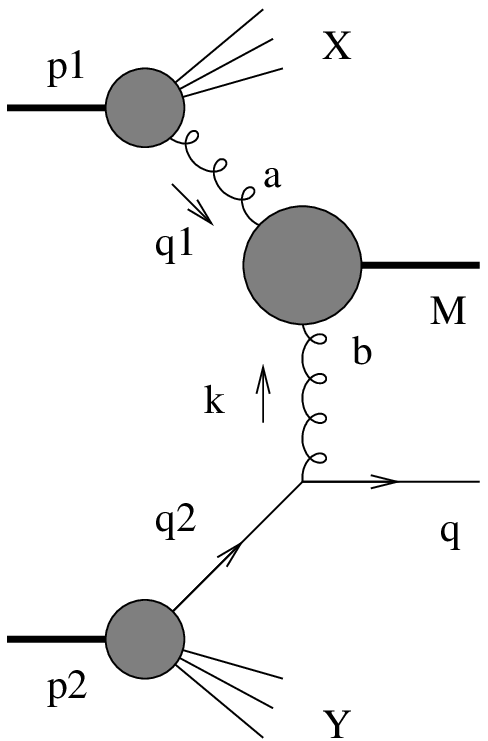}} \hspace{2cm} 
\psfrag{p1}{$\hspace{-.5cm} H(p_1)$}
\psfrag{p2}{$\hspace{-.5cm} H(p_2)$}
\raisebox{0cm}{\includegraphics[scale=.84]{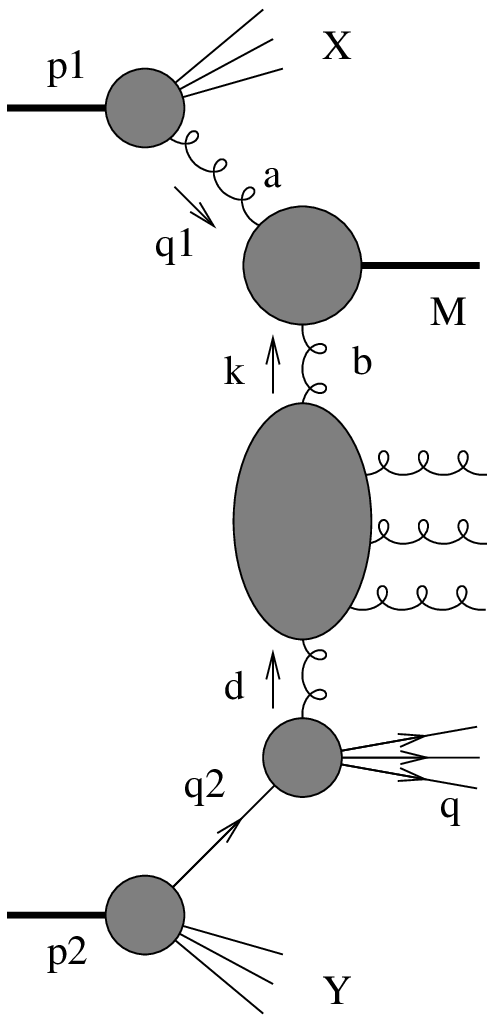}}} 
}
\caption{The high-energy hadroproduction of a meson $M$ and a jet (here originating from a quark) with a large rapidity difference between them. Left: Born approximation. Right: inclusion of BFKL-like resummation effects due to multiple emissions of gluons and of higher order jet vertex corrections.}
\label{Fig:Process}
\end{figure}

In order to reveal the so-called BFKL~\cite{Fadin:1975cb-Kuraev:1976ge-Kuraev:1977fs-Balitsky:1978ic} resummation effects specific to QCD in the perturbative Regge limit, first with 
leading logarithmic (LL) precision, and more recently at 
next-to-leading logarithmic (NLL) accuracy, the inclusive production of a dijet with
a large rapidity separation~\cite{Mueller:1986ey} is one of the most promising processes.
Recent full NLL $k_t$-factorization studies of these Mueller-Navelet  jets~\cite{Colferai:2010wu,Ducloue:2013hia,Ducloue:2013bva,Ducloue:2014koa-Caporale:2012ih-Caporale:2013uva-Caporale:2014gpa-Celiberto:2015yba} could successfully describe these events at the LHC~\cite{Khachatryan:2016udy}, exhibiting the very first signs of BFKL resummation effects at the LHC. 
We here report on a study~\cite{Boussarie:2017oae} where we apply a similar formalism to study the production of a forward $J/\psi$ meson and a very backward jet with a rapidity interval that is large enough to probe the BFKL dynamics but small enough for
both the $J/\psi$ and the jet to be in the detector acceptance at LHC experiments such as ATLAS or CMS.
 In this work the $J/\psi$ meson and the tagged jet 
 are produced by the interaction of two partons,
each collinear to one of the hadrons, resumming the effect of any number of accompanying unobserved partons, as usual in the $k_t$-factorization approach, see figure~\ref{Fig:Process}. We compare two different approaches for the description of $J/\psi$ production: the NRQCD formalism~\cite{Bodwin:1994jh}, and the Color Evaporation Model (CEM)~\cite{Fritzsch:1977ay-Halzen:1977rs}.

\section{Theoretical description}

The  inclusive high-energy hadroproduction process of a $J/\psi$, via two gluon fusion, with a large rapidity difference between the jet and the meson, in scattering of a hadron $H(p_1)$ with a hadron $H(p_2)$, is illustrated in figure~\ref{Fig:Process}. 
First, the longitudinal momentum fractions of the 
jet and of the meson are assumed to be large enough so that the usual collinear factorization applies, with hard scales provided by the heavy meson mass  and by the transverse momentum of the jet. We denote the momentum of the upper (resp. lower) parton as $x \, p_1$ (resp. $x' \, p_2$), their distribution being given by usual parton distribution functions (PDFs). Second, the $t-$channel exchanged momenta  between the meson and the jet cannot be neglected due to their large relative rapidity, calling for the use of $k_t-$factorization. 
In this picture,
the differential cross section reads
\begin{equation}
\frac{d\sigma}{dy_V d|p_{V\bot}|d\phi_V  dy_{J} d|p_{J\bot}|d\phi_{J}} 
= \sum_{{a, b}} \int_0^1 \!dx \int_0^1 \!dx' f_{a}(x) f_{b}(x') \frac{ d \hat{\sigma}}{dy_V d|p_{V\bot}|d\phi_V  dy_{J} d|p_{J\bot}|d\phi_{J}} \, ,
\end{equation}
where $f_{a, b}$ are the standard PDFs of a parton $a (b)$ in the according hadron and 
\begin{equation}
\frac{ d \hat{\sigma}}{dy_V d|p_{V\bot}|d\phi_V  dy_{J} d|p_{J\bot}|d\phi_{J}}
= \int d^2 k_\perp  \, d^2 k_\perp' V_{V,a}(k_\perp,x) \, G(-k_\perp,-k_\perp',\shat)\,V_{J, b}(-k_\perp',x') \, ,
\label{eq:bfklpartonic}
\end{equation}
being the partonic cross section in $k_t$-factorization, where 
 $G$ is the BFKL Green's function depending on $\shat=x x' s$. Here
 $\sqrt{s}$ is the center-of-mass energy  of the two colliding hadrons.

At leading order (LO), the jet is made of a single parton, and 
the  expressions for the jet vertex can be found in refs.~\cite{Bartels:2001ge-Bartels:2002yj}.
At next-to-leading order (NLO), the jet can be made of either a single or two partons. The explicit form of these jet vertices can be found in ref.~\cite{Colferai:2010wu} as extracted from refs.~\cite{Bartels:2001ge-Bartels:2002yj}, and
the LL~\cite{Fadin:1975cb-Kuraev:1976ge-Kuraev:1977fs-Balitsky:1978ic} and  NLL~\cite{Fadin:1998py-Ciafaloni:1998gs} BFKL Green's function $G$ can be found in ref.~\cite{Colferai:2010wu}. 

To determine the $J/\psi$ production vertex\footnote{In practice we only calculate the LO $J/\psi$ production vertex.}, we fix the normalization by focusing on the Born approximation (lhs of figure~\ref{Fig:Process}). We generically denote with an index $M$ the kinematical variables attached to the system made of the meson and the possible accompanying unobserved particles, and use an index $V$ for the kinematical variables attached to the $J/\psi$ meson itself. Denoting  ${\cal A}_\nu^{ab}$  the $S$-matrix element describing the $g g \to M$ transition, the differential cross-section reads
\begin{align}
\frac{d \sigma}{dy_{J}d|p_{J\bot}|d\phi_{J}}=\int & \, dx \, g(x)\, dx' \,H^q(x')\,d^2k_\bot\,\delta(x-[\alpha_M])\,\delta^2(k_\bot - [p_{M\bot}]) \left[  \frac{d^3p_M}{(2\pi )^32E_M} \right] \nonumber \\
& \times \frac{8\sqrt{2}\pi^2}{s^2(N^2-1)^2 \,x\,k^2_\bot}  \,  \sum_{[M]}  {\cal A}_{\mu \bot}^{ab} g^{\mu \nu}_\bot   ({\cal A}^{ab}_{\nu\bot})^*  \,\,V_{J,q}^{(0)}(-k_\bot,x')\,, \label{CSecM}
\end{align} 
in which we factorized out the vertex for quark jet production in the Born approximation,
\begin{equation}
V_{J,q}^{(0)}(r_\bot,x') = \frac{g^2}{4\pi \sqrt{2}} \frac{C_F}{|r_\bot|} \,\delta(1-\frac{x_J}{x'}) \,\delta^2(r_\bot - p_{J\bot})\,.
\label{Vq}
\end{equation}

\subsection{Color-singlet NRQCD contribution}

In the color singlet contribution, 
the system $[M]$ is made of the produced $J/\psi$ charmonium and of the unobserved gluon produced simultaneously with the charmonium in gluon-gluon fusion. 
The momenta $p_V$ of the $J/\psi$ and  $l$
of this gluon are parameterized in terms of Sudakov variables:
\begin{equation}
p_V=\alpha_V p_1 + \frac{M^2_{J/\psi}-p^2_{V\bot}}{\alpha_V s}p_2 +p_{V\bot}\;,\;\;\;\;\;l=\alpha_l \, p_1 - \frac{l_\bot^2}{\alpha_l s} p_2+l_\bot \,.
\label{CSSudakov}
\end{equation}
Integrating over  $l\,,$ the $J/\psi$ production vertex of the color singlet NRQCD contribution is 
\begin{equation}
V_{J/\psi}^{(1)} =\frac{|p_{V\bot}|\sqrt{2} }{2^5 \pi^4 s^2 (N^2-1)^2 k_\bot^2\, x} \frac{\theta(x-\alpha_V)}{x-\alpha_V} \sum_{\lambda_V, \lambda_l }  {\cal A}_{\mu \bot}^{ab} g_\bot^{\mu \nu}({\cal A}_{\nu \bot}^{ab})^* \,.
\label{CSVertex}
\end{equation}
The vertex to pass from open $q \bar{q}$ production, perturbatively calculable, to $J/\psi$ production  reads~\cite{Guberina:1980dc-Baier:1983va}
\begin{equation}
[v(q)\bar u(q)]^{ij}_{\alpha \beta} \rightarrow \frac{\delta^{ij}}{4N} \left(  \frac{\langle  {\cal O}_1 \rangle_V}{m} \right)^{1/2} \left[ \hat \epsilon^*_V \left(  2\hat q +2m \right)\right]_{\alpha \beta} ,
\label{CSvertex}
\end{equation}
with the momentum $q = \frac{1}{2}p_V$, $m$ being the mass of the charm quark, $M_{J/\psi}=2m$. Defining the non-perturbative coefficient 
$C_1 \equiv \sqrt{\langle  {\cal O}_1 \rangle_V/m},$
our result for the $J/\psi$ production vertex reads
\begin{equation}
V_{J/\psi}^{(1)} 
= \frac{|p_{V\bot}|\sqrt{2 } g^6 C_1^2}{s^2 \pi^4 2^{13} k_\bot^2} \frac{d^{abl} d^{abl}}{N^2(N^2-1)^2} \frac{\theta(x-\alpha_V)}{x(x-\alpha_V)}{\cal D}^{(1)}(J/\psi) \, ,
\label{vertexsin}
\end{equation}
with $\alpha_V=\frac{\sqrt{4m^2-p_{V\bot}^2}}{\sqrt{s}} e^{y_V}$ and $d^{abl} d^{abl} = \frac{(N^2-4)(N^2-1)}{N}$. See ref.~\cite{Boussarie:2017oae} for the  expression of 
${\cal D}^{(1)}(J/\psi)$.

\subsection{Color-octet NRQCD contribution}

In the color-octet contribution $[M]$ denotes the $J/\psi$ state alone. The $J/\psi$ production vertex is
\begin{equation}
V_{J/\psi}^{(8)}(k_\bot,x) =  \frac{|p_{V\bot}|\delta(x-\alpha_V)  \delta^2(k_\bot - p_{V\bot})  }{\sqrt{2} \pi s^2 (N^2-1)^2 \, k_\bot^2\, x} 
 \sum_{\lambda_{V}}  {\cal A}_{\mu \bot}^{ab} g_\bot^{\mu \nu}({\cal A}_{\nu \bot}^{ab})^* \, .
\label{VertexJPsiOctet }
\end{equation}
The vertex which allows to pass from open $q \bar{q}$ production to $J/\psi$ production is defined as
\begin{equation}
[v(q)\bar u(q)]^{ij\rightarrow d}_{\alpha \beta} \rightarrow\,t^d_{ij}  d_8\,
 \left( \frac{\langle{\cal O}_8 \rangle_V}{m} \right)^{1/2}\left[ \hat \epsilon^*_V \left(  2\hat q +2m \right)\right]_{\alpha \beta} \, ,
\label{COvertex}
\end{equation}
with $d_8=\frac{1}{4\sqrt{3}}.$
Our final result for the 
$J/\psi$ production vertex then reads
\begin{equation}
V_{J/\psi}^{(8)} =
 -\delta (x-\alpha_V) \delta^2(k_\bot - p_{V \bot}) \frac{|p_{V \bot}|\sqrt{2} g^4 k_\bot^2 x }{128\pi m^3(4m^2-k_\bot^2)^2} \langle{\cal O}_8 \rangle_V \,.
\label{Voctet}
\end{equation}

\subsection{Color evaporation model}

In the color evaporation model (COM) $[M]$ denotes an open quark-antiquark produced state with an invariant mass $M$ integrated in the range $[2m,2M_D]$. 
The $J/\psi$ production vertex reads
\begin{align}
V^{(\rm CEM)}_{J/\psi}(k_\bot,x) = F_{J/\psi}\,\int\limits_{4m^2}^{4M_{D}^2} & dM^2 \, \delta\left(M^2- \frac{m^2 -l_\bot^2}{\alpha \bar \alpha}\right)\frac{d\alpha\, d^2l_\bot}{\alpha \bar \alpha} \sum_{\lambda_{k_1}\lambda_{k_2}}  {\cal A}_{i \bot}^{ab} g_\bot^{i j}({\cal A}_{j \bot}^{ab})^* \nonumber \\
& \times \frac{|p_{V\bot}|\sqrt{2} \,\delta\left(x-\alpha_V) \delta^2(k_\bot - p_{V\bot}\right)}{2^5 \pi^4 s^2 (N^2-1)^2\, k^2_\bot\, x}  \,, \label{CEVertex}
\end{align}
with
\begin{align}
\sum_{\lambda_{k_1}\lambda_{k_2}}     ({\cal A}^{ab}_{i\bot})^*   g_{\bot }^{ij}   {\cal A}^{ab}_{j\bot} = \frac{g^4}{4} \left(   c_a \Tr_a + c_b \Tr_b  \right) \, .
\label{CEhard}
\end{align}
We refer to ref.~\cite{Boussarie:2017oae} for the expressions of
$\Tr_a$ and $\Tr_b$  and
of the two color structures $c_a$ and $c_b.$

\section{Results}

\def\figscale{0.74}

\psfrag{sigma}{\raisebox{1.5mm}{\scalebox{0.8}{$\displaystyle \frac{d\sigma}{d|p_{V \bot}|\, d|p_{J \bot}| \, dY} [{\rm nb.GeV}^{-2}]$}}}
\psfrag{cos}{\scalebox{0.9}{$\langle \cos \varphi \rangle$}}
\psfrag{Y}{\scalebox{0.9}{$Y$}}

\psfrag{2jets}{\scalebox{0.8}{2 jets}}
\psfrag{CS}{\scalebox{0.8}{Color singlet}}
\psfrag{CO}{\scalebox{0.8}{Color octet}}
\psfrag{CEM}{\scalebox{0.8}{Color evaporation}}

\def\factscale{.86}
\begin{figure}[h!]
\scalebox{\factscale}{
\hspace{0.5cm}\includegraphics[scale=\figscale]{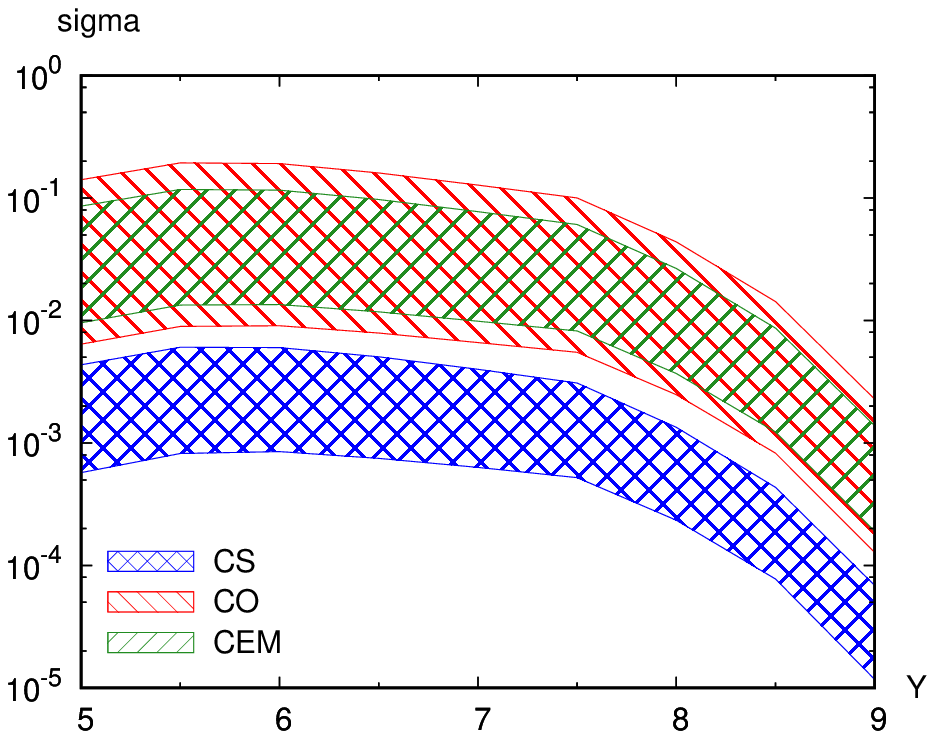}
\hspace{0.8cm}\includegraphics[scale=\figscale]{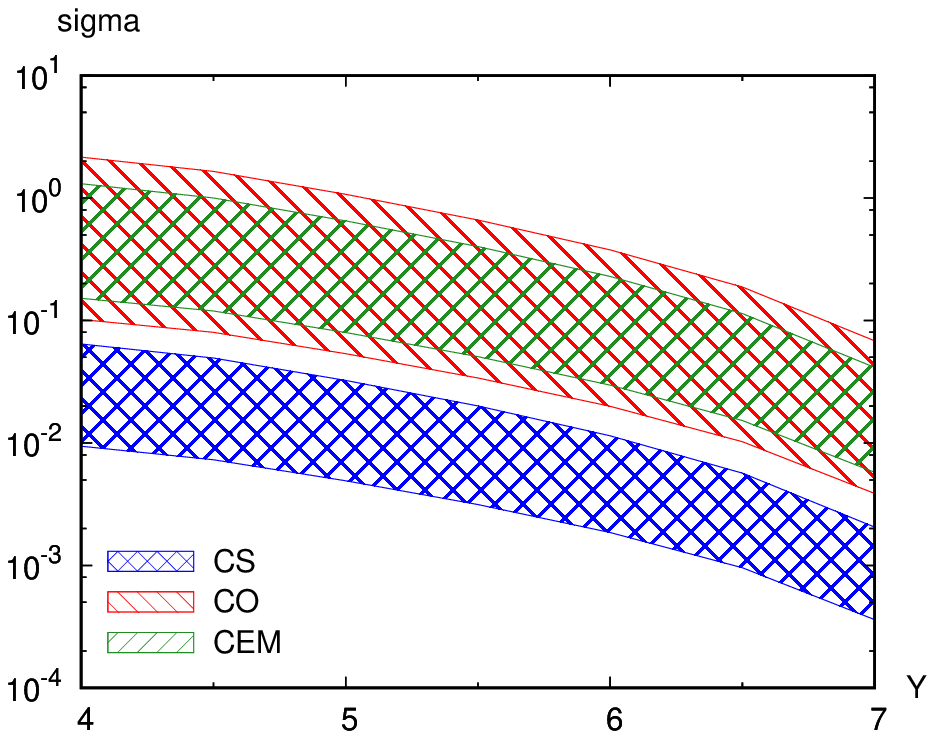}}

\hspace*{0.4cm} {\small $0<y_V<2.5, \; -6.5<y_J<-5, \; p_\bot=10$ GeV} \hspace{0.5cm} {\small $0<y_V<2.5, \; -4.5<y_J<0, \; p_\bot=10$ GeV}

\vspace{0.4cm}

\scalebox{\factscale}{\hspace{0.5cm}\includegraphics[scale=\figscale]{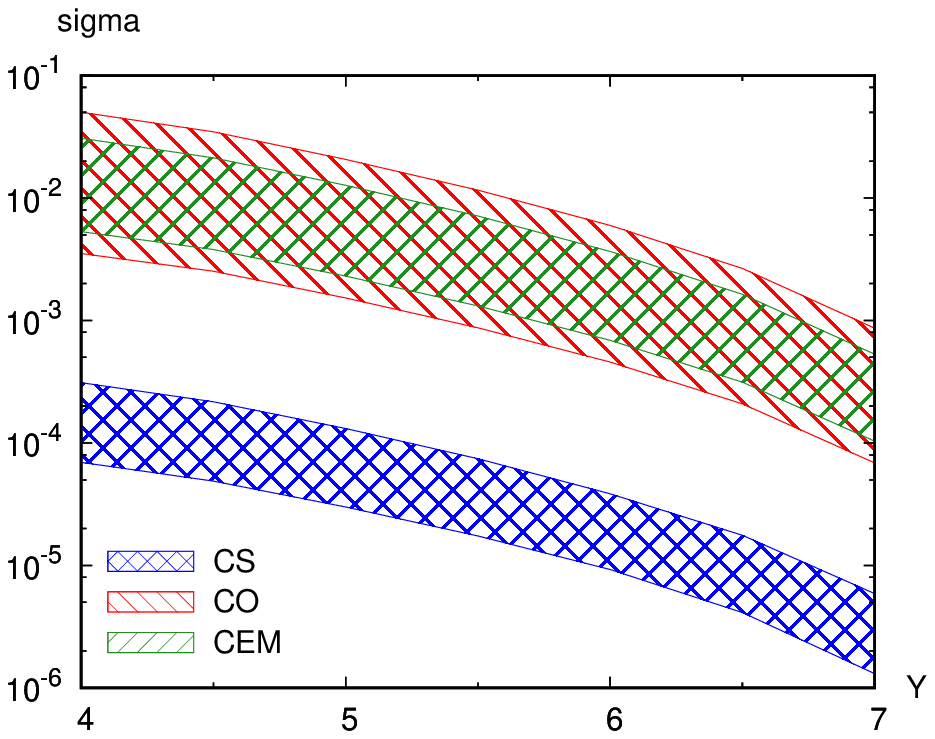}
\hspace{0.8cm}\includegraphics[scale=\figscale]{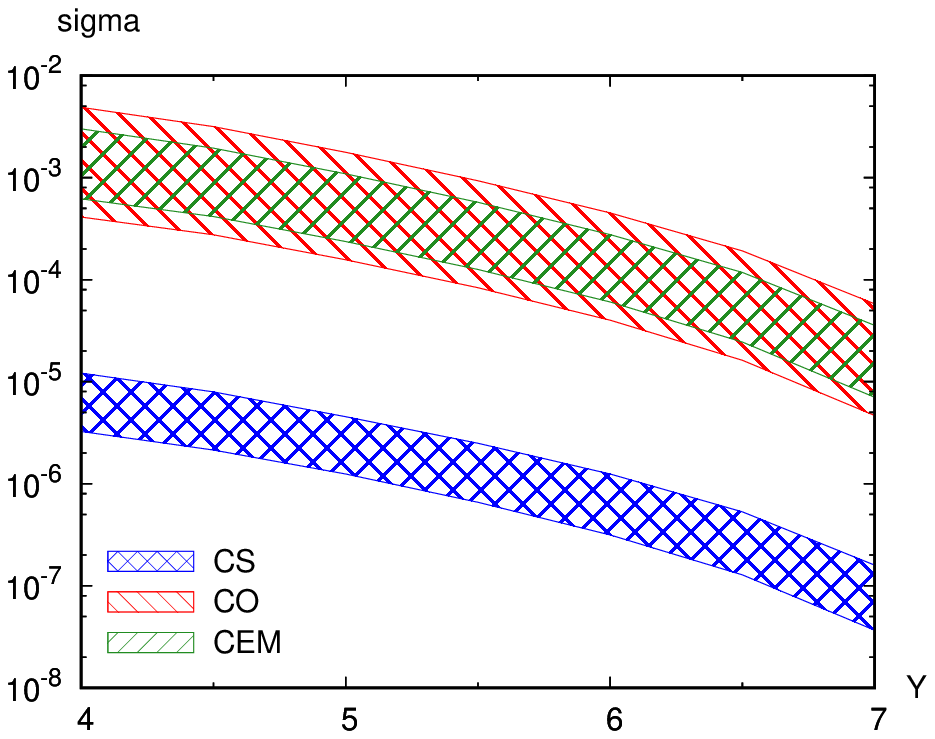}}

\hspace*{0.4cm} {\small $0<y_V<2.5, \; -4.5<y_J<0, \; p_\bot=20$ GeV} \hspace{0.8cm} {\small $0<y_V<2.5, \; -4.5<y_J<0, \; p_\bot=30$ GeV}
\caption{Cross section at $\sqrt{s}=13$ TeV as a function of the relative rapidity $Y$ between the $J/\psi$ and the jet, in four different kinematical configurations.}
\label{Fig:sigma-13}
\end{figure}
\begin{figure}[htbp]
\scalebox{\factscale}{\hspace{0.5cm}\includegraphics[scale=\figscale]{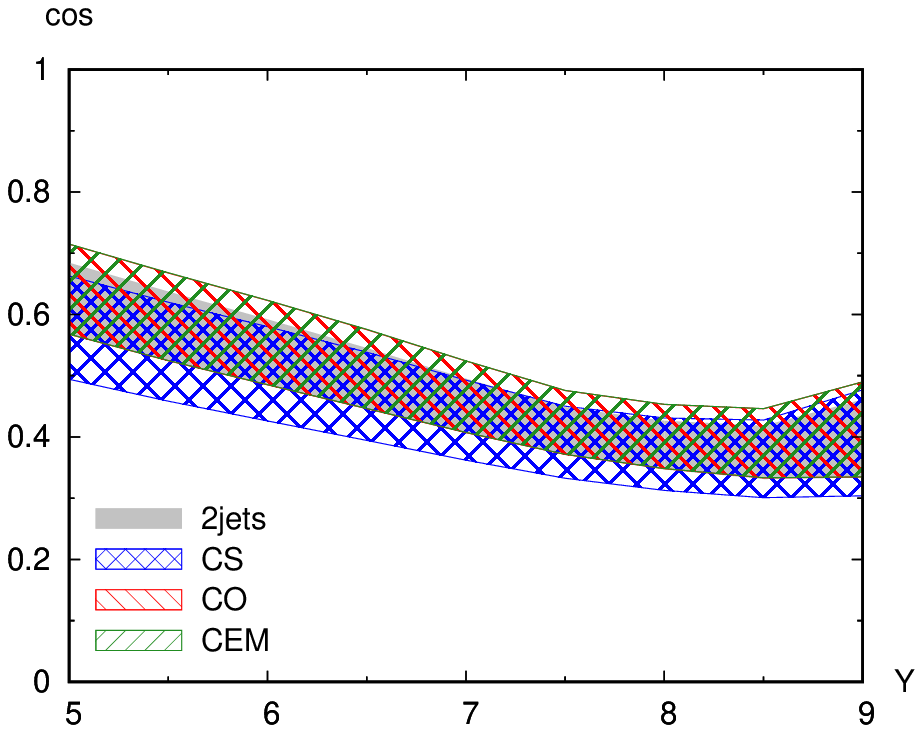}
\hspace{0.8cm}
\includegraphics[scale=\figscale]{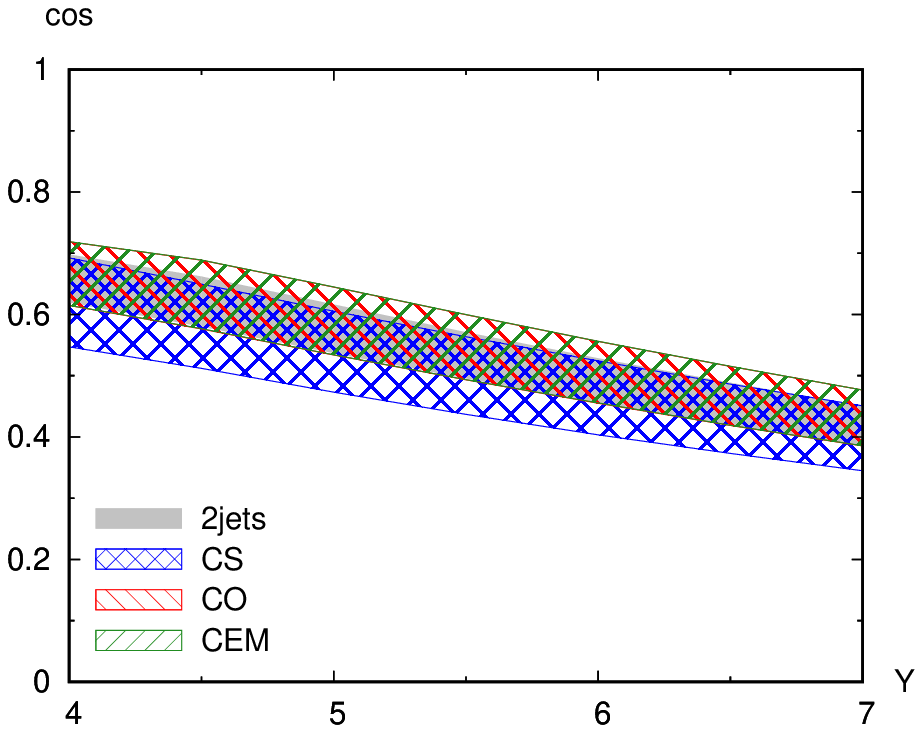}}

\hspace*{0.4cm} {\small $0<y_V<2.5, \; -6.5<y_J<-5, \; p_\bot=10$ GeV} \hspace{0.5cm} {\small $0<y_V<2.5, \; -4.5<y_J<0, \; p_\bot=10$ GeV}

\vspace{0.1cm}

\scalebox{\factscale}{\hspace{0.5cm}\includegraphics[scale=\figscale]{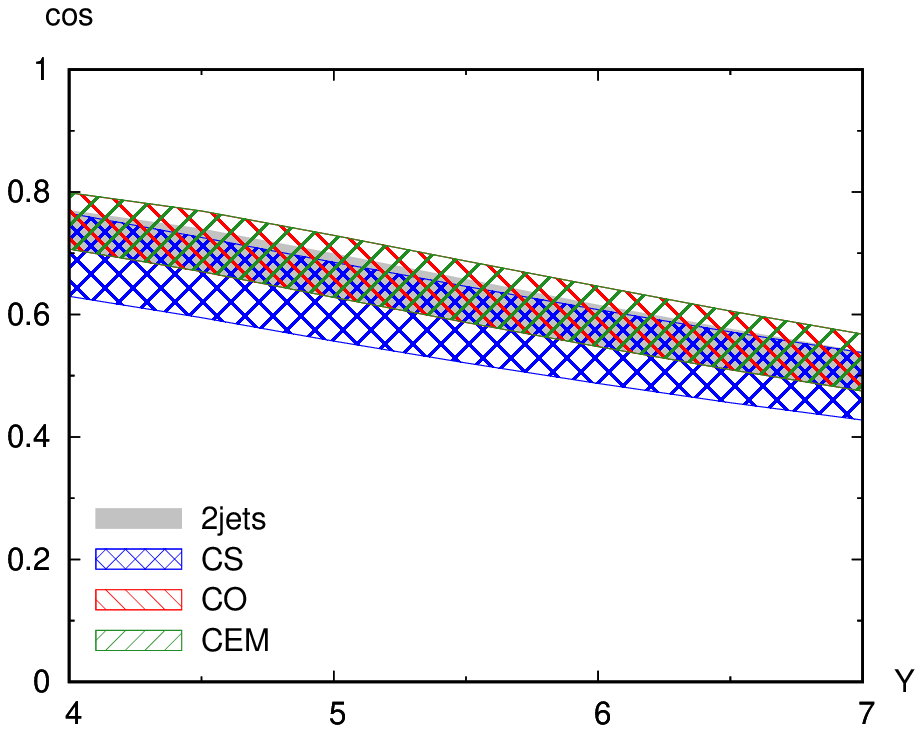}
\hspace{0.8cm}
\includegraphics[scale=\figscale]{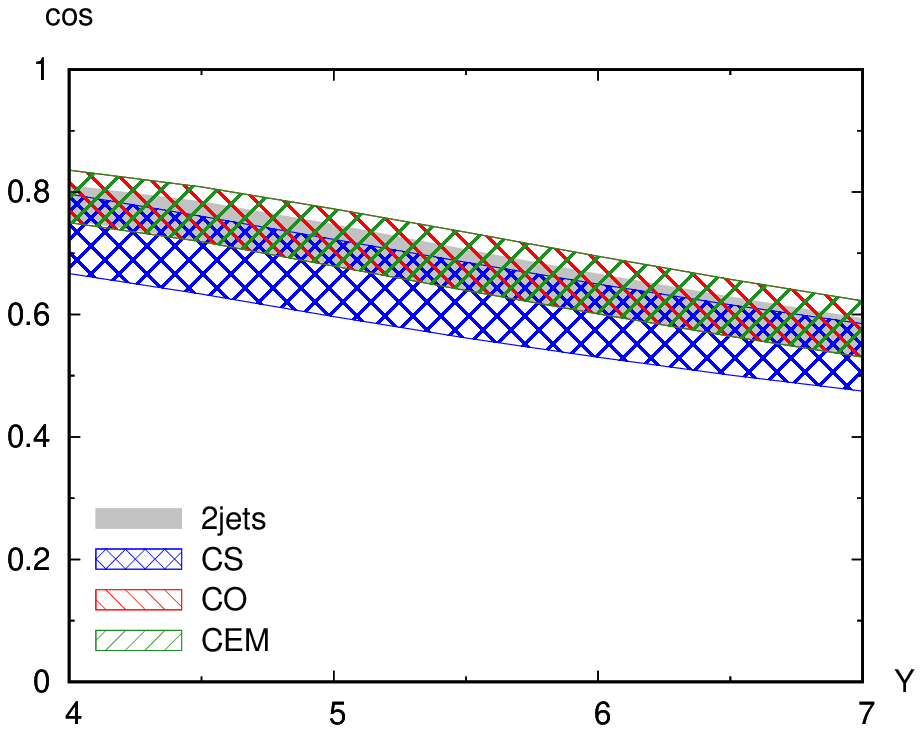}}

\hspace*{0.4cm} {\small $0<y_V<2.5, \; -4.5<y_J<0, \; p_\bot=20$ GeV} \hspace{0.8cm} {\small $0<y_V<2.5, \; -4.5<y_J<0, \; p_\bot=30$ GeV}
\caption{Variation at $\sqrt{s}=13$ TeV of $\langle \cos \varphi \rangle$ with $\varphi$ defined as $\varphi=|\phi_V-\phi_J-\pi|$,  as a function of the relative rapidity $Y$ between the $J/\psi$ and the jet, for the four kinematical cuts described in the text. The grey band corresponds to the results obtained when the $J/\psi$ production vertex is replaced by the leading order jet production vertex.}
\label{Fig:cos-13}
\end{figure}
We now compare the cross sections and azimuthal correlations between the $J/\psi$ meson and the jet obtained with the color singlet, color octet and color evaporation hadronization mechanisms, for $\sqrt{s}=13$ TeV (see ref.~\cite{Boussarie:2017oae} for the results at $\sqrt{s}=8$ TeV). We consider equal values of the transverse momenta of the $J/\psi$ and the jet, $|p_{V \bot}|=|p_{J \bot}|=p_\bot$, and four different kinematical configurations:
\begin{eqnarray}
\label{conf1}
&\bullet& 0<y_V<2.5, \; -6.5<y_J<5, \; p_\bot=10~{\rm GeV},\\
\label{conf2}
&\bullet& 0<y_V<2.5, \; -4.5<y_J<0, \; p_\bot=10~{\rm GeV},\\
\label{conf3}
&\bullet& 0<y_V<2.5, \; -4.5<y_J<0, \; p_\bot=20~{\rm GeV},\\
\label{conf4}
&\bullet& 0<y_V<2.5, \; -4.5<y_J<0, \; p_\bot=30~{\rm GeV}.
\end{eqnarray}
The  first configuration could be measured combining the CASTOR detector to tag the jet and the CMS tracking system to measure the $J/\psi$ meson. For the other three configurations the restriction in  rapidity corresponds to the typical values accessible by the main detectors at ATLAS and CMS. We only show results for $Y>4,$
where the
 BFKL approximation is valid.
We use the BLM renormalization scale fixing procedure, see refs.~\cite{Ducloue:2013bva} and \cite{Boussarie:2017oae} for details.
The uncertainty band is computed in the same way as in ref.~\cite{Ducloue:2013bva} 
with the addition of the variation of the non-perturbative constants $\langle{\cal O}_1 \rangle_V,$ $\langle{\cal O}_8 \rangle_V$
and $F_{J/\psi}$ in the ranges specified in ref.~\cite{Boussarie:2017oae}.
We fix the charm quark mass to $m=1.5$ GeV.
Our results for the differential cross-section and for the azimuthal correlations are shown respectively in figures~\ref{Fig:sigma-13} and \ref{Fig:cos-13}. We thus see that
CO model  and CEM lead to similar effects for the cross-section, and all models give the same prediction for $\langle \cos \varphi \rangle$.
Note that we did not include any double parton scattering contribution. For Mueller-Navelet jets, this  is expected to be rather small with respect to the single BFKL ladder contribution~\cite{Ducloue:2015jba}, except potentially for large $s$ and small jet transverse momenta, e.g. for the configuration (\ref{conf1}).

\paragraph*{Acknowledgements.}
\noindent
The research by R.B. and L.Sz. was supported by the National Science Center, Poland, grant No. 2015/17/B/ST2/01838. The research by B.D. was supported by the Academy of Finland, project 273464, and by the European Research Council, grant ERC-2015-CoG-681707.
This work is partially supported by
the French grant ANR PARTONS (Grant No. ANR-12-MONU-0008-01), by the COPIN-IN2P3 agreement,
by the Labex P2IO and by the Polish-French collaboration agreement
Polonium.
This work used computing resources from CSC -- IT Center for Science in Espoo, Finland.

\providecommand{\href}[2]{#2}\begingroup\raggedright\endgroup

\end{document}